\documentclass[12pt]{article}
\let\section=\subsection     \let\subsection=\subsubsection                
\usepackage{graphicx}

\begin{document}
\begin{center}
   {\large \bf The effect of early chemical freeze out on radial and}\\[2mm]
   {\large \bf elliptic flow from a full 3D hydrodynamic model}\\[5mm]
   T.~HIRANO$^a$ and K.~TSUDA$^b$ \\[5mm]
   {\small \it  $^a$Physics Department, University of Tokyo, Tokyo 113-0033, Japan \\}
   {\small \it  $^b$Department of Physics, Waseda University, Tokyo 169-8555, Japan \\[8mm] }
\end{center}

\begin{abstract}\noindent
We investigate the effect of early chemical freeze-out on radial and elliptic flow by using a fully three dimensional hydrodynamic model. We find that the time evolution of temperature and the thermal freeze-out temperature dependence of average radial flow are different from the results by using a conventional hydrodynamic model in which chemical equilibrium is always assumed. We also analyse the $p_t$ spectrum and $v_2(p_t)$ at the RHIC energy and consistently reproduce experimental data by choosing the thermal freeze-out temperature $T_{\mathrm{th}} = 140$ MeV.

\end{abstract}

\section{Introduction}
In the conventional hydrodynamic models for analyses of relativistic heavy-ion collisions, both chemical and thermal equilibrium are assumed and the chemical freeze-out temperature $T_{\mathrm{ch}}$ is the same as the thermal freeze-out temperature $T_{\mathrm{th}}$ which is so chosen as to reproduce the slope of $p_t$ (or $m_t$) spectra. For analyses of the $p_t$ slope at the RHIC energy by using the conventional hydrodynamics, see, e.g., Refs.~[1--5].
On the other hand, various analyses based on thermal models indicate the following picture of time evolution of hot and dense hadronic matter \cite{REVIEW}: the system first undergoes the chemical freeze-out at $T_{\mathrm{ch}}$ where the observed particle ratios are fixed, continues to cool down and finally goes through the thermal freeze-out where the momentum distribution is fixed.\footnote{The \textit{observed} momentum distribution can be affected by feeding from resonance decay after thermal freeze-out.}
We take account of this picture in a hydrodynamic model and discuss how the early chemical freeze-out affects the space-time evolution of fluids. We also analyse the particle spectrum at the RHIC energy. In the present work, we assume $\mu_B = 0$ for simplicity. Although we neglect finite baryon effects, our hydrodynamic simulations are performed \textit{in a genuine three dimensional space}.
As a future work, we will include finite $\mu_{\mathrm{B}}$ in our model and analyse particle ratios and spectra at various collision energies \cite{ATTEMPT,ATTEMPT2}.

\section{Equations of State}
We construct three models equation of state (EOS) to discuss the space-time evolution of fluids. These models describe the first order phase transition between the QGP phase and the hadron phase at $T_{\mathrm{c}} = 170$ MeV. We suppose the QGP phase is composed of massless u, d, s quarks and gluons and that it is common to three models. For the hadron phase, we choose three different models EOS as follows.

The first model is a conventional resonance gas model in which complete chemical equilibrium is always assumed (model CE). We include hadrons up to the mass of $\Delta(1232)$ for simplicity.

The second model is the simplest one which describes the picture of chemical freeze-out (model CFO). Below $T_{\mathrm{ch}}$, we assume the numbers of \textit{all} hadrons $N_i$ included in the EOS are fixed and that the particle number densities obey $\partial_\mu n_i^\mu = 0$. We introduce a chemical potential $\mu_i(T)$ associated with each species so that $N_i$ becomes a conserved quantity. From the conservation of entropy $\partial_\mu s^\mu = 0$, the ratio of the particle number density to the entropy density obeys
\begin{eqnarray}
\label{RATIO1}
\frac{n_i(T<T_{\mathrm{ch}})}{s(T<T_{\mathrm{ch}})} = \frac{n_i(T_{\mathrm{ch}})}{s(T_{\mathrm{ch}})}
\end{eqnarray}
for all hadrons along the adiabatic path.
Here we assume $T_{\mathrm{ch}} = 170$ MeV, which is consistent with a recent analysis based on a thermal model \cite{PBM}.
From Eq.~(\ref{RATIO1}), we obtain a chemical potential as a function of temperature for each hadron.
\begin{center}
   \includegraphics[width=12cm,height=4cm]{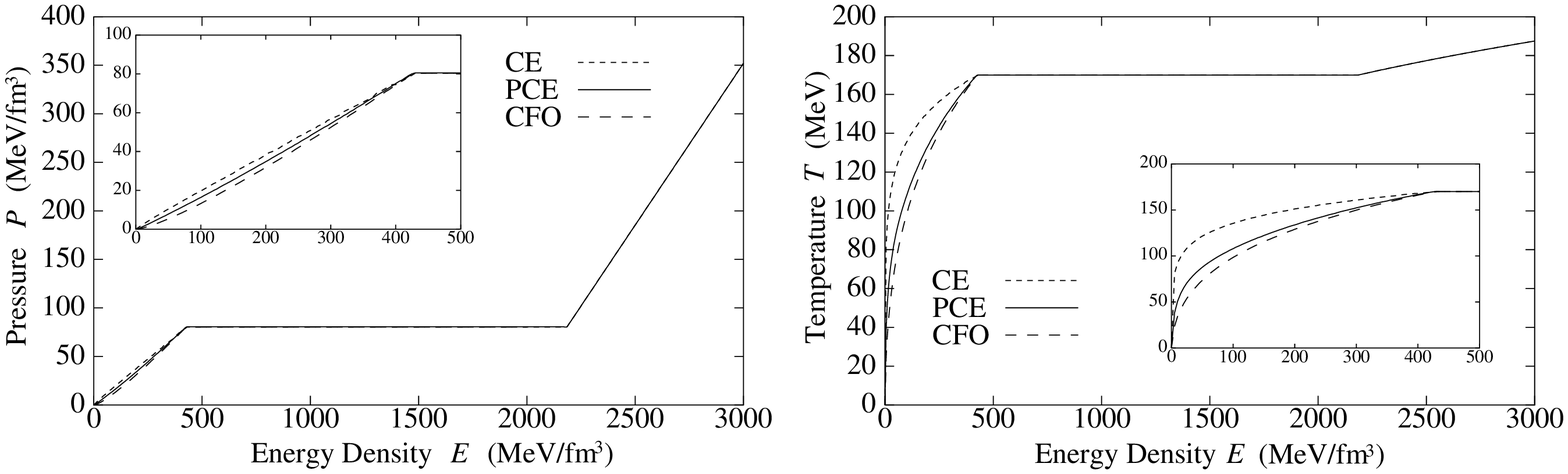}\\
   \parbox{14cm}
        {\footnotesize 
        Fig.~1: (Left) Pressure as functions of energy density for three models (Right) Temperature as functions of energy density.}
\end{center}

The third model represents a more realistic EOS than the second one. The observed particle numbers are always composed of the direct particles and the contribution from resonance decays, e.g., $\bar{N}_\pi = N_\pi + \sum_{i\neq \pi} d_{i \rightarrow \pi} N_i$. So some processes with large cross sections can be equilibrated even below $T_{\mathrm{ch}}$ as long as the equality $\bar{n}_i(T<T_{\mathrm{ch}})/s(T<T_{\mathrm{ch}}) = \bar{n}_i(T_{\mathrm{ch}})/s(T_{\mathrm{ch}})$ is kept instead of Eq.~(\ref{RATIO1}). We assume $\pi$, $K$, $\eta$, $N$, $\Lambda$ and $\Sigma$ are ``stable" particles and that all chemical potentials can be represented by chemical potentials associated with these stable particles, e.g., $\mu_\rho = 2\mu_\pi$, $\mu_{K^*} = \mu_\pi + \mu_K$, $\mu_\Delta = \mu_\pi + \mu_N$ and so on \cite{BEBIE}. Thus the third model describes the \textit{partial} chemical equilibrium (model PCE) even below $T_{\mathrm{ch}}$.

One can easily evaluate EOS's for these models by using chemical potentials as obtained above. We represent pressure and temperature as functions of energy density for three models in Fig.~1. We find pressure as a function of energy density is similar to each other. On the other hand, temperature as a function of energy density in the models CFO or PCE is deviated from the model CE. See also Fig.~1 (right). Since the resonance population of the models CFO or PCE is larger than that of the model CE due to the chemical freeze-out, the energy density at a fixed temperature in the hadron phase is also large in those models. 
Conversely, temperature in the models CFO or PCE at a fixed energy density is \textit{smaller} than in the model CE. 

\section{Radial and Elliptic Flow}

By using the EOS's in the previous section, we perform the hydrodynamic simulation with appropriate initial conditions and compare the hydrodynamic evolution with each other. We choose initial conditions so as to roughly reproduce pseudorapidity distribution in Au+Au 130$A$ GeV central (0-6 \%) and semi-central (35-45 \%) collisions obtained by the PHOBOS Collaboration \cite{PHOBOS}. At the initial time $\tau_0=0.6$ fm/$c$, the energy density (temperature) for central collisions at the centre of the fluid, where it has a maximum value, is 30.8 GeV/fm$^3$ (349.5 MeV). For further details of our hydrodynamic model, especially initial longitudinal and transverse profiles, see Ref.~\cite{HIRANO1}.

\begin{center}
   \includegraphics[width=12cm,height=4cm]{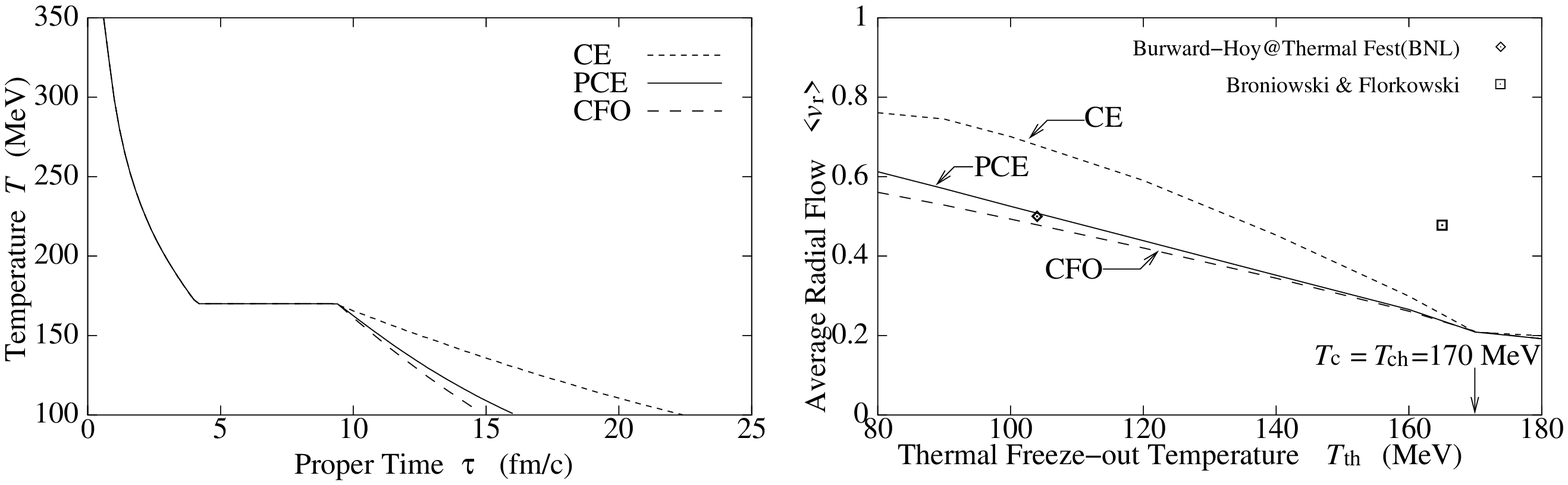}\\
   \parbox{14cm}
        {\footnotesize 
        Fig.~2: (Left) Time evolution of temperature for three models. (Right) Thermal freeze-out temperature dependence of average radial flow for three models.}
\end{center}

Figure 2 (left) shows the time evolution of temperature at the centre of the fluid for three models. From EOS's represented in Fig.~1 (left), we expect the time evolution of \textit{energy density} is similar to each other.
However, when we transpose the energy density to the temperature by using EOS's in Fig.~1 (right), the difference between three models appears in the time evolution of \textit{temperature}. We find temperature in the models CFO or PCE cools down more rapidly than in the conventional model CE. 
We next show in Fig.~2 (right) the thermal freeze-out temperature dependence of average radial flow at vanishing space-time rapidity $\eta_s =0$ and compare our results with the ones from thermal models.
Radial flows in the models CFO and PCE are considerably reduced.
This is understood qualitatively by the same reason as the above discussion on the time evolution of temperature.
These results seem to be consistent with the one obtained by Burward-Hoy \cite{BURWARD-HOY}, while these are completely inconsistent with the model I$\!$I in Ref.~\cite{BRONIOWSKI}. 
Since the results from thermal models are different from each other and we do not take into account the baryonic chemical potential, more detailed analyses based on both thermal and hydrodynamical models are needed to understand the \textit{thermal} freeze-out.

\begin{center}
   \includegraphics[width=12cm,height=4cm]{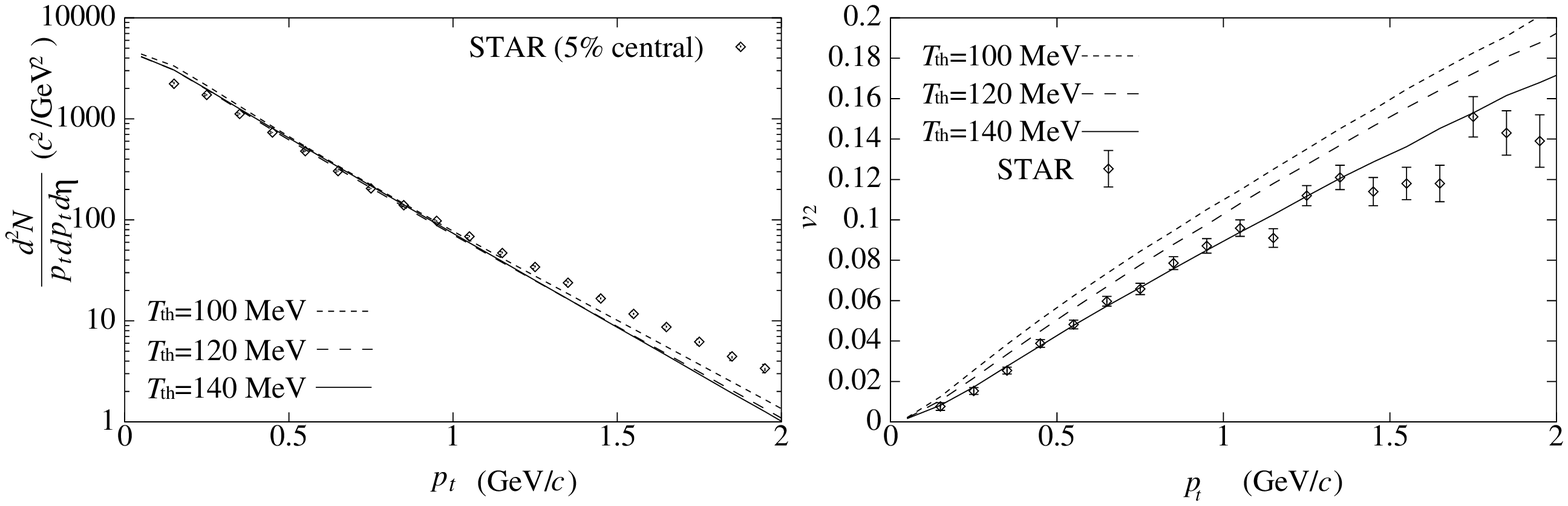}\\
   \parbox{14cm}
        {
        \footnotesize 
        Fig.~3: Thermal freeze-out temperature dependence of transverse momentum spectrum for negative charged hadrons (left) and $v_2(p_t)$ for charged hadrons (right). Solid, dashed and dotted lines correspond to, respectively, $T_{\mathrm{th}}$ = 140, 120 and 100 MeV}
\end{center}

Finally we show in Fig.~3 the $p_t$ spectrum and $v_2(p_t)$ only for the model PCE. Left figure represents the transverse momentum distribution of negative charged hadrons for various thermal freeze-out temperatures. Experimental data is observed by the STAR Collaboration \cite{STAR.PT}.
Although the $p_t$ slope usually depends on $T_{\mathrm{th}}$ in the conventional EOS where the chemical equilibrium is assumed,
our result for the model PCE is found to be almost independent of $T_\mathrm{th}$ due to the reduction of radial flow. Figure 3 (right) shows the transverse momentum dependence of $v_2$ for charged hadrons in minimum bias events. Our result with $T_\mathrm{th} = 140$ MeV gives the best fit to the data obtained by the STAR Collaboration \cite{STAR.V2} below 1 GeV/$c$. The discrepancy of $T_{\mathrm{th}}$ between our result and the one in Ref.~\cite{BURWARD-HOY} may be due to the fact that the resonance decays and the chemical potential of pions are not taken into account in Ref.~\cite{BURWARD-HOY}. 
We consistently reproduce experimental data of both $p_t$ spectrum and $v_2(p_t)$ below 1 GeV/$c$ by choosing $T_{\mathrm{th}} = 140$ MeV, so contribution from hard processes can become dominated above 1 GeV/$c$.
Note that the concavity of $v_2(p_t)$ in small $p_t$ region ($p_t< 0.3$ GeV/$c$) is due to the resonance decay after thermal freeze-out \cite{HIRANO2}.

\section{Summary and Discussion}
We have investigated the effect of early chemical freeze-out on radial and elliptic flow at the RHIC energy. We found the system cools down more rapidly than the conventional model due to the large population of resonances. We also found that the resultant radial flow is considerably reduced when we compare with each other at a fixed $T_{\mathrm{th}}$.
Although the space-time evolution of temperature in the model PCE is considerably different from the conventional model, we consistently reproduced $p_t$ spectrum and $v_2(p_t)$ below 1 GeV/$c$ by choosing the thermal freeze-out temperature $T_{\mathrm{th}} = 140$ MeV.
This indicates the contribution from hard processes can become dominated above 1 GeV/$c$. 
It should be noted that the average radial flow results in $<v_r>\mid_{\eta_{\mathrm{s}}=0} \sim 0.35 c$.

Since the two pion correlation function is believed to be sensitive to the size and the lifetime of the fluid, it is interesting to see the effect of early chemical freeze-out on the HBT radii.  Early chemical freeze-out in hydrodynamics will be one of the possible explanations to interpret the ``HBT puzzle" \cite{HEINZ} at the RHIC energy \cite{STAR.HBT,PHENIX.HBT}. In addition to the HBT radii, detailed analyses of particle ratio and spectra with finite $\mu_{\mathrm{B}}$ are also our future works.

\end{document}